\documentclass{article}

\usepackage{PRIMEarxiv}

\usepackage[utf8]{inputenc} 
\usepackage[T1]{fontenc}    
\usepackage{hyperref}       
\usepackage{url}            
\usepackage{booktabs}       
\usepackage{amsfonts}       
\usepackage{nicefrac}       
\usepackage{microtype}      
\usepackage{lipsum}
\usepackage{fancyhdr}       
\usepackage{graphicx}       
\graphicspath{{media/}}     

\usepackage{amsmath}
\usepackage{amssymb}
\usepackage{multirow}
\usepackage{algorithm}
\usepackage{algorithmic}
\usepackage{color}

\title{IMPORTANT-Net: Integrated MRI Multi-Parameter Reinforcement Fusion Generator with Attention Network for Synthesizing Absent Data
}

\author{
  Tianyu Zhang, Tao Tan\thanks{\textit{Corresponding Author}: Tao Tan (taotanjs@gmail.com)}, Luyi Han, Xin Wang, Yuan Gao, Jonas Teuwen, Regina Beets-Tan, Ritse Mann \\
  Department of Radiology \\
  The Netherlands Cancer Institute (NKI) \\
  Amsterdam \\
  \texttt{\{t.zhang, t.tan, l.han, x.wang, y.gao, j.teuwen, r.beetstan, r.mann\}@nki.nl} \\
}

\begin{document}
\maketitle

\begin{abstract}
Magnetic resonance imaging (MRI) is highly sensitive for lesion detection in the breasts. Sequences obtained with different settings can capture the specific characteristics of lesions. Such multi-parameter MRI information has been shown to improve radiologist performance in lesion classification, as well as improving the performance of artificial intelligence models in various tasks. However, obtaining multi-parameter MRI makes the examination costly in both financial and time perspectives, and there may be safety concerns for special populations, thus making acquisition of the full spectrum of MRI sequences less durable. 
In this study, different than naive input fusion or feature concatenation from existing MRI parameters, a novel \textbf{I}ntegrated MRI \textbf{M}ulti-\textbf{P}arameter reinf\textbf{O}rcement fusion generato\textbf{R} wi\textbf{T}h \textbf{A}tte\textbf{NT}ion Network (IMPORTANT-Net) is developed to generate missing parameters. First, the parameter reconstruction module is used to encode and restore the existing MRI parameters to obtain the corresponding latent representation information at any scale level. Then the multi-parameter fusion with attention module enables the interaction of the encoded information from different parameters through a set of algorithmic strategies, and applies different weights to the information through the attention mechanism after information fusion to obtain refined representation information. Finally, a reinforcement fusion scheme embedded in a $V^{-}$-shape generation module is used to combine the hierarchical representations to generate the missing MRI parameter. Results showed that our IMPORTANT-Net is capable of generating missing MRI parameters and outperforms comparable state-of-the-art networks.
Our code is available at \url{https://github.com/Netherlands-Cancer-Institute/MRI_IMPORTANT_NET}.
\end{abstract}


\section{Introduction}
\label{sec:introduction}
Breast cancer is the most common cancer in women worldwide and the leading cause of cancer death in women~\cite{sung2021global}. Early screening, diagnosis and treatment of breast cancer are of great significance for increasing the survival period of breast cancer patients~\cite{goldhirsch2013personalizing}. In breast imaging, breast magnetic resonance imaging (MRI) is a highly sensitive imaging modality for breast cancer detection, characterization and diagnosis. Different sequences/parameters of MRI can capture the specific features of breast lesions. For example, repeated T1-weighted imaging before and after contrast administration can delineate enhancement abnormalities, T2-weighted imaging with fat suppression can easily visualize cysts, T2-weighted imaging without fat suppression can better delineate lesion morphology, and diffusion-weighted imaging (DWI) can provide information on cell density and tissue microstructure based on the diffusion of tissue water~\cite{mann2019breast,van2021factors}. Some studies have shown that multiparametric MRI could improve the performance of radiologists' diagnosis of lesions, as well as the performance of artificial intelligence models in certain tasks, including classification, diagnosis, and segmentation of breast lesions~\cite{dalmis2019artificial,hu2020deep,conte2021generative}. However, obtaining multi-parametric MRI makes the examination costly from financial and time considerations, thus making acquisition of an exhaustive multiparametric MRI protocol less optimal~\cite{dar2019image,zhou2020hi}. Currently, MRI-based breast cancer screening is highly valued by clinicians and researchers~\cite{gommers2021breast,sanderink2021diffusion,geuzinge2021cost,veenhuizen2021supplemental,mann2022breast}. Typically, only T1-weighted pre- and post- contrast MRI and DWI or T2 are included in the screening protocols, and while some abbreviated MRI protocols have also explored obtaining more MRI parameters, this is time-consuming and costly~\cite{mann2019breast,mango2015abbreviated,chen2017abbreviated,leithner2019abbreviated,kim2022abbreviated}. Unfortunately, this may lead to missing characteristic information of unscanned MRI sequences, such as T2-weighted MRI, that may sometimes be valuable for breast lesion characterization~\cite{heacock2016evaluation,mann2019breast,naranjo2022mri}. However, since there is an overlap in the physical properties of MRI sequences, a large part of the missing parameters might be deducted from the available sequences. Therefore, a technique that can synthesize missing/absent MRI parameters from existing ones would be of great value.

With the continuous development of computer technology, artificial intelligence-based methods have shown potential in data generation and image transfer, and generative adversarial networks (GAN) have received extensive attention due to their potential ability to generate data and enable image-to-image translation through an unsupervised game of generator and discriminator~\cite{goodfellow2020generative}. Subsequently, a large number of GAN-based works have emerged~\cite{creswell2018generative,karras2019style,gui2021review}. In medical imaging, rapid adoption of such networks, with many traditional and novel applications, such as image reconstruction, segmentation, detection, classification, and cross-modality synthesis~\cite{yi2019generative} are being explored. Among them, cross-modality synthesis is considered as a potentially effective method to generate desired but absent modalities from the existing ones~\cite{yi2019generative,zhu2017unpaired,isola2017image,zhou2020hi,li2022virtual}. However, these methods can only perform mutual synthesis between two modalities and can not utilize all existing data, and even with advanced modelling~\cite{zhou2020hi,li2022virtual}, the performance is still unsatisfactory.In addition, the existing methods do not effectively fuse multi-parameter/multi-modal information, and do not reflect the special contribution of different parameters/modalities to the synthesized target image.

In this study, we developed a novel model, called \textbf{I}ntegrated MRI \textbf{M}ulti-\textbf{P}arameter reinf\textbf{O}rcement fusion generato\textbf{R} wi\textbf{T}h \textbf{A}tte\textbf{NT}ion Network (IMPORTANT-Net). First, the existing parameters are encoded and reconstructed by the reconstruction module to obtain latent representation information, then the fusion module interacts and fuses the encoded multi-parameter features. An inter-parameter attention module is designed to obtain refined features, and finally the hierarchical refined representation information is used to synthesize missing/absent MRI sequences. Furthermore, we also validate the effectiveness of IMPORTANT-Net in synthesizing missing/absent data on another publicly available brain MRI dataset from Brain Tumor Segmentation 2021 (BraTS2021)~\cite{menze2014multimodal,baid2021rsna}. Our contributions are as follows:
\begin{itemize}
    \item[i] The multi-parameter fusion module is designed through a special set of algorithmic strategies to enable the interaction between different parameters from the parameter reconstruction module.
    \item[ii] The multi-parameter representation attention module is designed to optimize the special contributions of different parameters.
    \item[iii] The hierarchical reinforcement fusion in $V^{-}$-shape generation network achieves the synthesis of missing/absent data.
    \item[iv] Our IMPORTANT-Net outperforms state-of-the-art methods, with an ablation study on the input level demonstrating the incremental value from multiple parameters in image synthesis.
\end{itemize}

\section{Related Work}
\label{sec:relatedwork}
Recently, some studies have shown that some GAN-based models can effectively perform mutual synthesis between MR and CT, PET and CT, PET and MR, 3T and 7T MR~\cite{yi2019generative, armanious2020medgan}. In addition, related studies have also shown that GAN-based models can perform mutual synthesis between multi-parameters of MRI~\cite{welander2018generative,liu2019susan,dar2019image,yu2019ea,yurt2021mustgan}, such as T1-weighted and T2-weighted MRI.
Emami et al.~\cite{emami2018generating} developed a GAN model using a residual network as the generator and a CNN as the discriminator to generate CT using T1-weighted MRI images. However, the simple model structure cannot effectively extract the underlying representation information, and the application scenarios are limited, leading to the inability to utilize multi-modal images (here using only T1-weighted MRI as input), limiting the performance and generalizability of their models. Many previous works, such as Wolterink et al.~\cite{wolterink2017deep}, Jin et al.~\cite{jin2019deep}, and Dar et al.~\cite{dar2019image} used CycleGAN~\cite{zhu2017unpaired}-based networks to synthesize target images. However, the synthesis effect of Cycle-GAN-based methods is not ideal and cannot incorporate multimodal information. Maspero et al.~\cite{maspero2018dose} tried to synthesize target images based on pix2pix-GAN~\cite{isola2017image}. Although this method can combine multi-modal images at the input, it does not fuse the feature information of different modalities, so the model performance is not satisfactory. Recently, some works, such as Hi-Net~\cite{zhou2020hi} and MMgSN-Net~\cite{li2022virtual}, have attempted to use multi-modal images to synthesize absent images.
However, simple fusion or feature connection of existing images cannot effectively fuse multi-parameter/multi-modal information, thus limiting the application scenarios and performance of their models. Different than naive input fusion or feature concatenation from existing images, and to overcome the limitations of variable application scenarios, here we present a novel model exploiting reinforcement fusion of existing data to efficiently synthesize missing/absent data.
\begin{figure*}[htbp]
	\centering
	\includegraphics[width=1\linewidth]{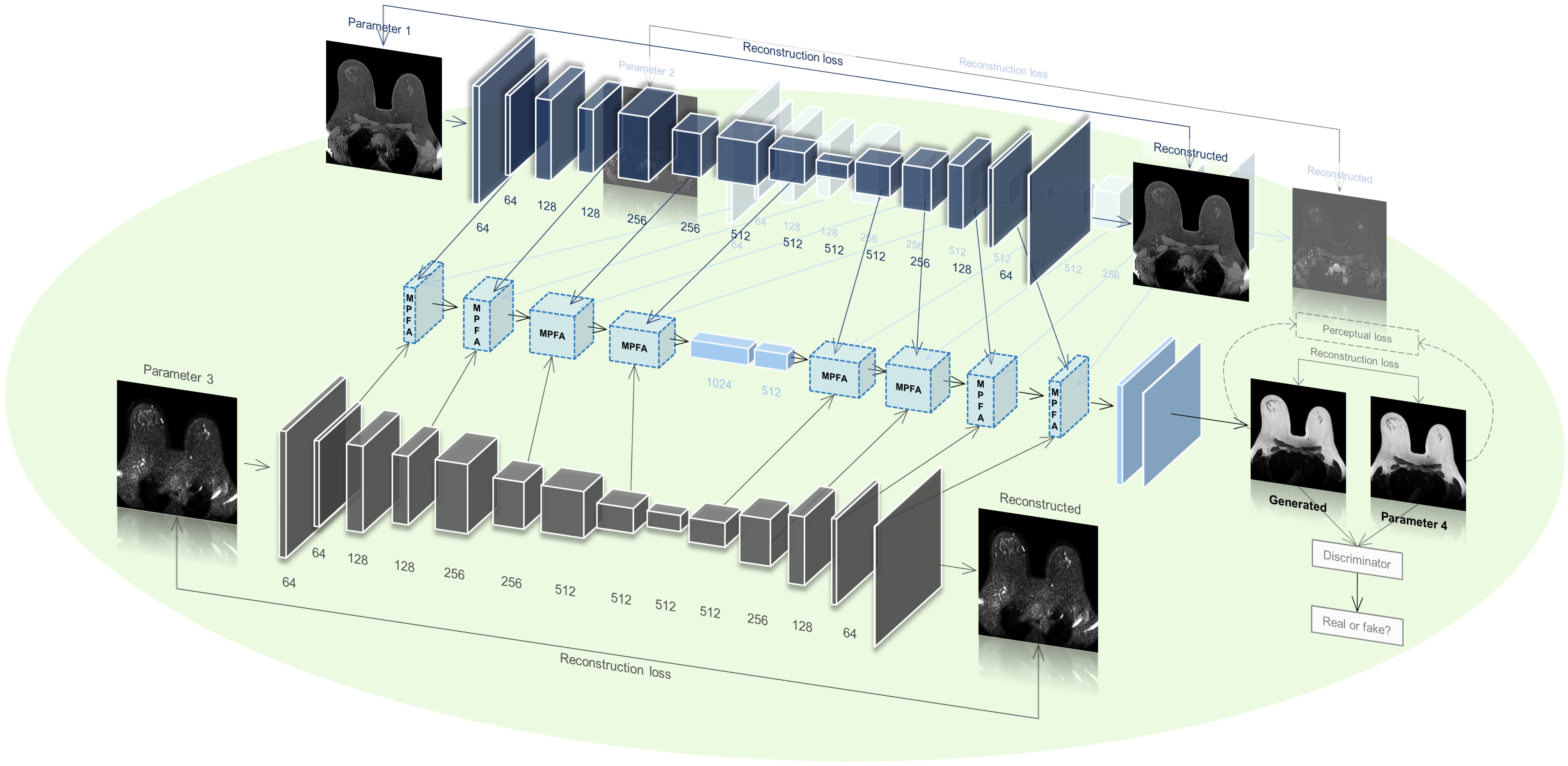}
	\caption{The structure of our proposed IMPORTANT-Net.}
        \label{fig:fig1}
	\vspace{-10pt}
\end{figure*}

\section{Approach}
\label{sec:method}
\subsection{Parameter reconstruction module}
To fully exploit multi-parameter information and mine latent representations, we propose an \textbf{I}ntegrated MRI \textbf{M}ulti-\textbf{P}arameter reinf\textbf{O}rcement fusion generato\textbf{R} wi\textbf{T}h \textbf{A}tte\textbf{NT}ion Network (IMPORTANT-Net), which integrates different functional modules with respective missions. The first functional module is called "parameter reconstruction module" (as shown in Fig.~\ref{fig:fig1}), which is used to automatically encode and reconstruct the information of each input MRI parameter, and mine the latent representations of different MRI parameters at the multi-scale level for subsequent multi-parameter information fusion. The reconstruction loss function is set as follows:
\begin{equation}
    \label{eq:reconstruction}
    \begin{aligned}
        \mathcal{L}_{Rec}(R_i)=&\sum_{i}^{n_p}{\|{p}_{i}-R_{i}({p}_{i})\|_1}
    \end{aligned}
\end{equation}
where $p_i$ represents individual parameter of MRI, $R_{i}({p}_{i})$=$f_{\theta_{i}^{re}}({p}_{i})$ represents the reconstructed image of $p_i$, and ${\theta_{i}^{re}}$ represents the corresponding network parameters of our parameter reconstructor $\mathcal{R}_{i}$.

\subsection{Multi-parameter fusion with attention module}
\label{sec:MPFA}
The second functional module is called "multi-parameter fusion with attention module (MPFA)", which is used to fuse the representation information from different parametric MRIs and make deep interaction between the information of different parameters. In MPFA module, to enforce explicit interaction, the MAPS strategy (element-wise Maximization, element-wise Average, element-wise Product, and element-wise Subtract) is employed to enable the interaction of information from different parameters. As shown in Fig.~\ref{fig:fig2}, the output $P_i^{dn}$ of the n-th down-sampling layer of the $i$-th MRI parameter or the $P_i^{um}$ of the $m$-th up-sampling layer of the $i$-th MRI parameter is obtained from the corresponding layer of the parameter reconstruction module, and then they are fed to the “interaction room” to achieve information fusion. Then, the features before and after the interaction are concatenated as $P_{concat}\in{\Bbb{R}^{{C}\times{H}\times{W}}}$. In particular, considering the different contributions of representations from different parametric MRIs, we design a parametric attention mechanism to impose different weights on features to obtain refined features. First, average pooling and max pooling are applied to $P_{concat}$ to obtain the corresponding ${Pool}_{ave}\in{\Bbb{R}^{C\times1\times1}}$ and ${Pool}_{max}\in{\Bbb{R}^{C\times1\times1}}$, respectively. Then, ${Pool}_{ave}$ and ${Pool}_{ave}$ are added element-wise after passing through a shared fully-connected ($f_{\theta^{fc}}$) neural network to obtain ${M}_{add}\in{\Bbb{R}^{C\times1\times1}}$. Then, the parameter-level attention map ${A}_{para}\in{\Bbb{R}^{C\times1\times1}}$ is obtained by nonlinear transformation of ${M}_{add}$. Finally, the refined feature $P_{concat}^{'}\in{\Bbb{R}^{{C}\times{H}\times{W}}}$ is obtained through the element-wise multiplication of $P_{concat}$ and ${A}_{para}$. The corresponding formula is as follows:
\begin{figure*}[htbp]
	\centering
	\includegraphics[width=1\linewidth]{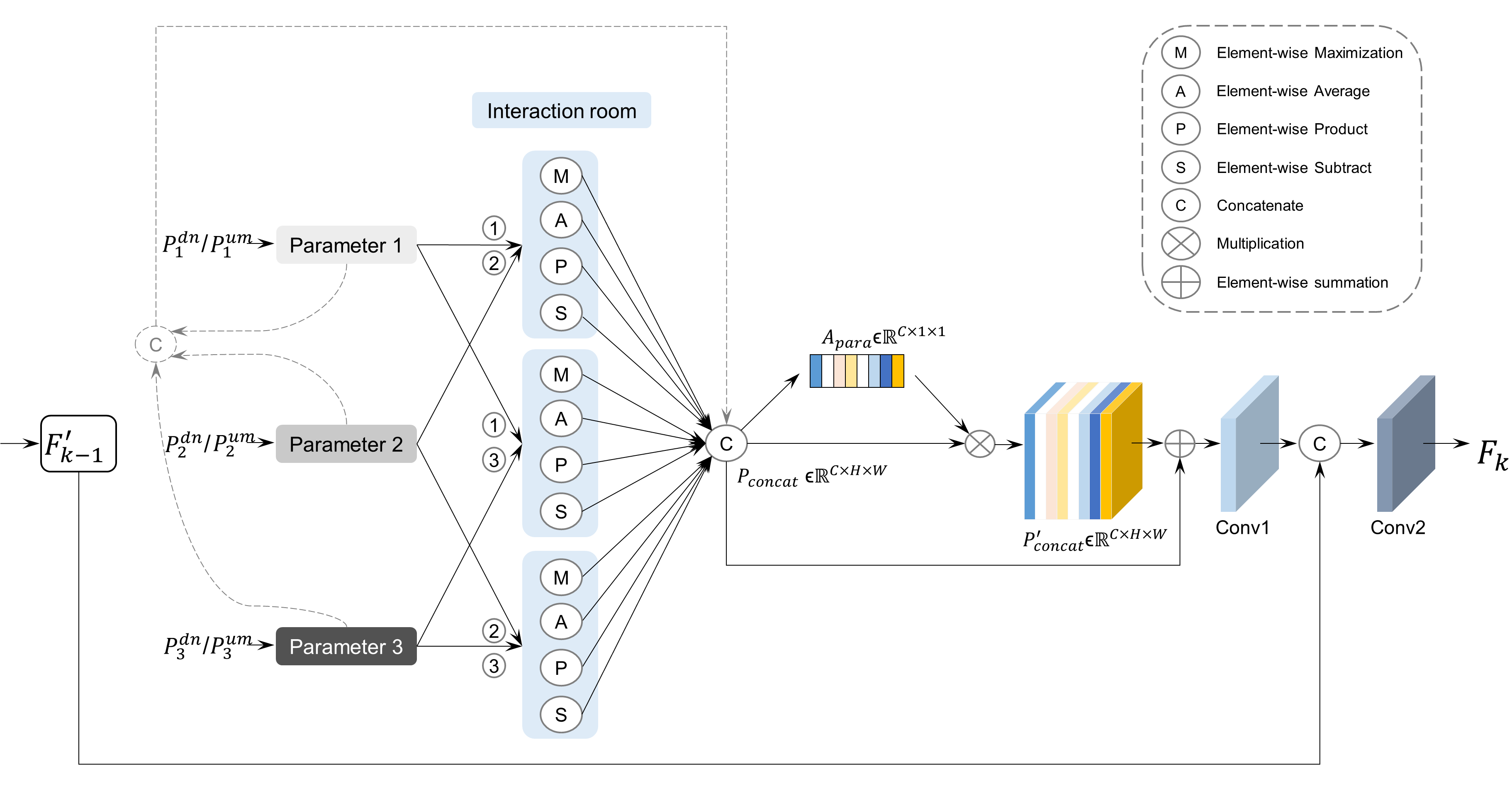}
	\caption{Details of the multi-parameter fusion with attention module.}
        \label{fig:fig2}
	\vspace{-10pt}
\end{figure*}

\begin{equation}
    \label{eq:mpfa}
    \begin{aligned}
        &{P_{concat}^{'}\in{\Bbb{R}^{{C}\times{H}\times{W}}}}={P_{concat}}\otimes{{A}_{para}}  \\
        &={P_{concat}}\otimes({\sigma({f_{\theta^{fc}}}({AvePool({P_{concat}})}}\oplus{f_{\theta^{fc}}}({MaxPool}({P_{concat}})))))
    \end{aligned}
\end{equation}
where $\otimes$ represents element-wise multiplication, $\oplus$ represents element-wise summation, $\sigma$ represents the sigmoid function, $\theta^{fc}$ represents the corresponding network parameters of the shared fully-connected neural network, and $AvePool$ and $MaxPool$ represent average pooling and maximum pooling operations, respectively. Subsequently, $P_{concat}$ and $P_{concat}^{'}$ are added element-wise and fed to the convolutional layer Conv1, whose output is concatenated with $F_{k-1}^{'}$ (features after down-sampling/up-sampling of the output $F_{k-1}$ of the $(k-1)$-th MPFA module), and then fed to Conv2, and finally the output $F_k$ of the $k$-th MPFA module is obtained.

\subsection{Hierarchical ${V^-}$-shape reinforcement generation module}
\begin{figure}[htbp]
	\centering
	\includegraphics[width=1\linewidth]{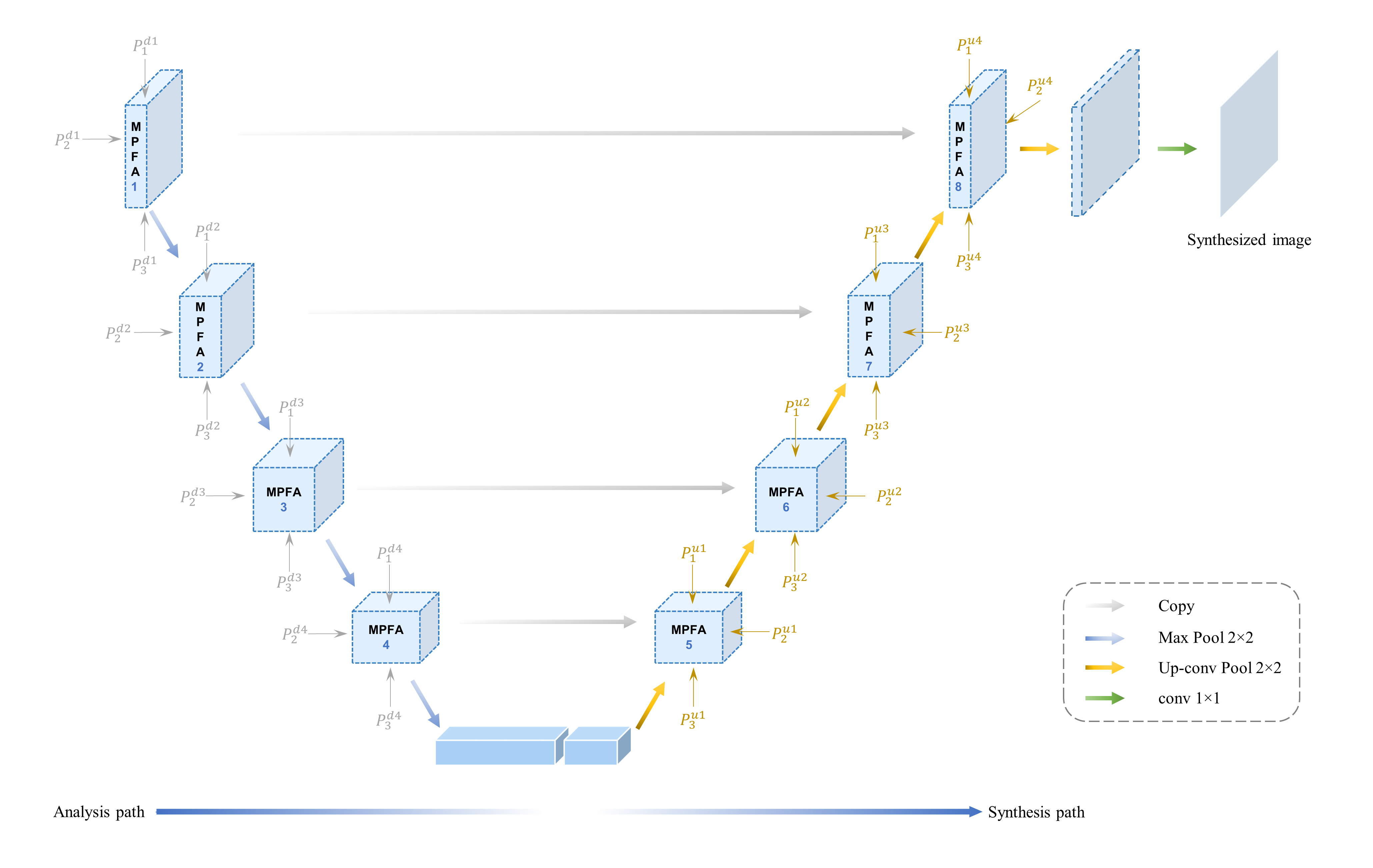}
	\caption{The structure of the hierarchical $V^-$-shape reinforcement generation module.}
    \label{fig:fig3}
\end{figure}

The third functional module is called the "hierarchical ${V^-}$-shape reinforcement generation module ", which is used to synthesize the target image based on the hierarchical structure and skip connections according to the hierarchical representation information. As shown in Fig.~\ref{fig:fig1}, the "parameter reconstruction module" is responsible for mining the potential representations of each MRI parameter, and MPFA is responsible for fusing the representations of each MRI parameter. Finally, the hierarchical MPFA embedded in a $V^{-}$-shape generation module is responsible for global analysis and reinforcement fusion, and final synthesis of the target image. Fig.~\ref{fig:fig3} shows the details of this module. MPFA1-4 is called the "analysis path", where the MPFA fuses the information of each MRI parameter ($P_i^{dn}$), and is then fed to the next MPFA after maximum pooling. In particular, the features after the "analysis path" are sequentially fed to two convolutional layers with 1024 and 512 filters of size 3$\times$3, respectively, and the output is then sent to the next path. MPFA5-8 is called the "synthesis path", where MPFA combines the corresponding feature information in the previous analysis path while fusing the multi-parameter MRI representation $P_i^{um}$, and then is fed to the next MPFA after up-sampling. 
\vspace{-5pt}
\vspace{-10pt}

\subsection{Loss function}
In the synthesis process, the generator $\mathcal{G}$ tries to generate an image according to the input multi-parameter MRI ($p_1$, $p_2$, $p_3$), and the discriminator $\mathcal{D}$ tries to distinguish the generated image ${G}$($p_1$, $p_2$, $p_3$) from the real image $y$, and at the same time, the generator tries to generate a realistic image to mislead the discriminator. The generator's objective function is as follows:
\begin{equation}
    \label{eq:LG}
    \begin{aligned}
        \mathcal{L}_{G}(G,D)=&\Bbb{E}_{{p_1},{p_2},{p_3}{\sim}{{pro}_{data}({p_1},{p_2},{p_3})}}\\
        &[\log(1-{D({p_1},{p_2},{p_3},({G}({p_1},{p_2},{p_3})))})] \\
        &+\lambda_{1}\Bbb{E}_{{p_1},{p_2},{p_3},y}[{\|y-{G}({p_1},{p_2},{p_3})\|_1}]
    \end{aligned}
\end{equation}
and the discriminator's objective function is as follows:
\begin{equation}
    \label{eq:LD}
    \begin{aligned}
        \mathcal{L}_{D}(G,D)=&\Bbb{E}_{y{\sim}{{pro}_{data}(y)}}[\log{D(y)]}\\
        &+\Bbb{E}_{{p_1},{p_2},{p_3}{\sim}{{pro}_{data}({p_1},{p_2},{p_3})}}\\
        &[\log{(1-{D({G}({p_1},{p_2},{p_3}))})}]
    \end{aligned}
\end{equation}
where $pro_{data}(p_1,p_2,p_3)$ represents the empirical joint distribution of inputs $p_1$, $p_2$ and $p_3$ (multi-parameter MRI), $\lambda_1$ is a nonnegative trade-off parameter, and $l_1$-norm is used to measure the difference between the generated image and the corresponding ground truth.

In particular, the perceptual loss is introduced into our model as one of the objective functions to constrain the original and generated images at the depth feature level~\cite{johnson2016perceptual,zhang2018unreasonable}.  The deep features extracted by the neural network usually obtain deeper semantic information of the image with the deepening of the network layers. Therefore, the deep semantic information of the target image can be generated by penalizing the deep feature difference model. The objective function is as follows:
\vspace{-5pt}
\begin{equation}
    \label{eq:LP}
    \begin{aligned}
        \mathcal{L}_{P}(G)=&\sum_{layer=1}^{N} {\alpha_{layer}}\frac{1}{C_{j}H_{j}W_{j}}{\|{\phi_{j}}(y)-{\phi_{j}}({G}({p_1},{p_2},{p_3}))\|_1}
    \end{aligned}
\end{equation}
where $\phi$ represents the pre-trained network (VGG19 here), $j$ represents the $j$-th layer of the pre-trained network, $C_{j}H_{j}W_{j}$ represents the shape of the feature map of the $j$-th layer, $N$ represents the total number of selected layers, $l_1$-norm is used to measure the difference between the deep features of the generated image and the corresponding ground truth, and $\alpha_{layer}$ represents the weight of the corresponding selected layer. Here, all 5 max pooling layers of the VGG19 pre-trained model were selected to calculate the perceptual loss, so the corresponding formula is as follows:
\begin{equation}
    \label{eq:LP5}
    \begin{aligned}
        \mathcal{L}_{P}(G)= {\alpha_{1}}{\mathcal{L}_P^1} + {\alpha_{2}}{\mathcal{L}_P^2} + {\alpha_{3}}{\mathcal{L}_P^3} + {\alpha_{4}}{\mathcal{L}_P^4} + {\alpha_{5}}{\mathcal{L}_P^5}
    \end{aligned} 
\end{equation}
Therefore, the IMPORTANT-Net can be formulated with the following objective function:

\begin{equation}
    \label{eq:adversarial}
    \begin{aligned}
        \mathcal{L}(G,D,R_i)={\mathcal{L}_{G}(G,D)} + {\mathcal{L}_{D}(G,D)} + {\lambda_{2}}{\mathcal{L}_{Rec}(R_i)} + {\lambda_{3}}{\mathcal{L}_{P}(G)} 
    \end{aligned}
\end{equation}
where $\lambda_2$ and $\lambda_3$ are trade-off parameters.
We aim to solve:
\begin{equation}
    \label{eq:adv}
    \begin{aligned}
        {G}^*={\rm arg}\min_{G,R_i}\max_{D}\mathcal{L}(G,D,R_i)
    \end{aligned}
\end{equation}

\subsection{Visualization}
To visually explain the performance of models, visualizing the difference between the synthetic image $G(p_1,p_2,p_3)$ and the corresponding ground truth $y$ is needed to obtain qualitative intuition. Since changes on grayscale images cannot be well perceived by humans, we generate a color error map $H_{col}$ based on a given grayscale heatmap $H_{sub}$,
\begin{equation}
    \label{eq:viz}
    \begin{aligned}
        {H}_{col}=\psi(\omega({H}_{sub}))=\psi(\omega(|y-{G}({p_1},{p_2},{p_3})|))
    \end{aligned}
\end{equation}
where $\psi$ represents the application of the colormap on the given image. $\omega$ means to limit the absolute error in $H_{sub}$ to be between (0, $max$), where $max$ represents the displayed maximum value. After passing through $\psi$, the maximum value in $H_{sub}$ will be red, and the minimum value will be blue, resulting in the final colored error map $H_{col}$.

\subsection{Evaluation metrics}
Results analysis was performed by Python 3.7. Structural Similarity Index Measurement (SSIM), Peak Signal-to-Noise Ratio (PSNR), Normalized Mean Squared Error (NMSE) and perceptual loss $L_P$ were used as metrics to evaluate the synthetic performance, all formulas as follows:

\begin{equation}
    \label{eq:ssim}
    \begin{aligned}
        {SSIM}=\frac{(2\mu_{y(x)}\mu_{G(x)}+{c}_{1})(2\sigma_{y(x)G(x)}+{c}_{2})}{({\mu_{y(x)}^{2}}+{\mu_{G(x)}^{2}}+{c}_{1})({\sigma_{y(x)}^{2}}+{\sigma_{G(x)}^{2}}+{c}_{2})}
    \end{aligned}
\end{equation}

\begin{equation}
    \label{eq:psnr}
    \begin{aligned}
        {PSNR}=10\log_{10}\frac{\max^{2}(y(x), G(x))}{\frac{1}{N}\|y(x)-G(x)\|_{2}^{2}}
    \end{aligned}
\end{equation}

\begin{equation}
    \label{eq:nmse}
    \begin{aligned}
        {NMSE}=\frac{y(x)-G(x)}{\|y(x)\|_{2}^{2}}
    \end{aligned}
\end{equation}
where ${G}(x)$ represents a generated image, $y(x)$ represents a ground-truth image, $\mu_{y(x)}$  and $\mu_{{G}(x)}$  represent the mean of $y(x)$ and $G(x)$, respectively, $\sigma_{y(x)}$  and $\sigma_{G(x)}$  represent the variance of $y(x)$ and ${G}(x)$, respectively, $\sigma_{{y(x)}{G}(x)}$  represents the covariance of $y(x)$ and ${G}(x)$, and $c_1$ and $c_2$ represent positive constants used to avoid null denominators.
\vspace{-15pt}
\section{Experimental Results}
\label{sec:result}
\subsection{Data collection and Pre-processing}
In this study, an in-house dataset of 375 cases of T1-weighted, Dynamic Contrast-Enhanced (DCE), DWI and corresponding T2-weighted MRI were retrospectively collected from breast patients scanned at the Cancer Institute was created. First, all MRIs were resampled to 1 mm isotropic voxels and uniformly sized, resulting in volumes of 224$\times$224 pixel images with 150 slices per MRI. Then, the registration was performed based on Advanced Normalization Tools (ANTs)~\cite{avants2011reproducible}. We focus on Symmetric Normalization methods because of their proven reliability, speed and flexibility, which combines affine and deformable transformations, and uses mutual information as its optimization metric. During the registration process, the T1-weighted MRI of each case was considered as fixed volume, and all other parameters including DCE, DWI and T2-weighted MRI were considered as moving volumes. And then, after the corresponding deformation fields are obtained, the registered volumes are obtained by spatial transformation.
In addition, the BraTS2021 dataset~\cite{baid2021rsna, menze2014multimodal} was used for the extension study. This includes brain MRI of 1251 cases with four parameters for each case, including T1, T1-contrast-enhanced (T1ce), T2 and T2-fluid-attenuated inversion recovery (Flair). Images of size 180$\times$180 were cropped out from 2D slices, and then all MRIs were uniformly sized to volumes of 224$\times$224 pixel images with 155 slices.

\subsection{Experiment settings}
Based on the ratio of 8:2, the training set and independent test set of the in-house dataset have 300 and 75 cases, respectively, and the training set and independent test set of the BraST2021 dataset have 1001 and 250 cases, respectively. During training, the weights $\alpha_1$, $\alpha_2$, $\alpha_3$, $\alpha_4$, and $\alpha_5$ of the corresponding selected layers in the perceptual loss are set to 1/16, 1/8, 1/4, 1/2 and 1, respectively, and the trade-off parameters  $\lambda_3$,  $\lambda_1$ and  $\lambda_2$ were set to 200, 100 and 25, respectively. For IMPORTANT-Net, the batch was set to 16 for 200 epochs, the initial learning rate was 1e-3 with a decay factor of 0.9 every 5 epochs. $Adam$ optimizer was applied to update the model parameters. All models were trained on NVIDIA RTX A6000 graphics processing unit (GPU), 48 Gigabytes of GPU memory. Cycle-GAN~\cite{zhu2017unpaired}, pix2pix (P2P)-GAN~\cite{isola2017image}, Hi-Net~\cite{zhou2020hi} and MMgSN-Net~\cite{li2022virtual} were used as cohort models for synthesizing missing/absent data.

\subsection{Results comparison}
To evaluate our proposed model, we first calculated the results of synthetic T2-weighted breast MRI to obtain quantitative metrics and compared them with cohort models, including the now popular models such as Cycle-GAN, P2P-GAN, and Hi-Net. The quantitative indicators used include PSNR, SSIM, NMSE and $L_P$. As shown in Table~\ref{table1}, the SSIM of cycle-GAN and P2P-GAN models based on T1-weighted MRI to synthesize T2-weighted MRI is 83.71 $\pm$ 4.02 and 85.36 $\pm$ 2.84, respectively, and the results are worse than other models. This is due to the model limitation of cycle-GAN and P2P-GAN, they can only perform image synthesis from one MRI parameter to another MRI parameter (here based on T1-weighted MRI to synthesize T2-weighted MRI), but cannot combine multiple parameter MRI, and thus have to disregard important information from other parameters, resulting in poor results. Hi-Net and MMgSN-Net can combine MRI with two different parameters to synthesize missing image data. As shown in Table~\ref{table1}, the SSIM of Hi-Net and MMgSN-Net based on T1-weighted MRI and DWI to synthesize T2-weighted MRI is 87.64 $\pm$ 2.90 and 89.70 $\pm$ 2.95, respectively, which outperform cycle-GAN and P2P-GAN models. This indicates that combining multi-parameter MRI can improve the synthetic performance. However, the structure of the existing models still needs to be improved, and model innovation is needed to improve the synthesis performance. In contrast, the SSIM of IMPORTANT-Net based on T1-weighted MRI and DWI to synthesize T2-weighted MRI is 91.23 $\pm$ 3.09, outperforming all cohort models, which indicates that the images synthesized by IMPORTANT-Net have higher structural similarity to real images. Furthermore, the quantitative results are further improved after combining the three modalities using IMPORTANT-Net, with an SSIM of 91.82 $\pm$ 3.15 demonstrating the incremental value from multiple parameters in image synthesis.

\begin{table*}[htbp]
\centering
\caption{Results for synthesizing T2-weighted breast MRI for different models on the in-house dataset.}
\label{table1}
\setlength{\tabcolsep}{5pt}
\begin{tabular}{llcccc}
\hline 
Method & Parameter & SSIM$\uparrow$ & PSNR$\uparrow$ & NMSE$\downarrow$ & $L_{P}$$\downarrow$ \\ \hline
Cycle-GAN~\cite{zhu2017unpaired} & T1$\to$T2 &83.71 $\pm$ 4.02 & 23.59 $\pm$ 2.02 &0.1377 $\pm$ 0.042 &2.659 $\pm$ 0.501 \\
P2P-GAN~\cite{isola2017image} & T1$\to$T2 &85.36 $\pm$ 2.84 & 25.34 $\pm$ 1.09 &0.0815 $\pm$ 0.026 &1.903 $\pm$ 0.427 \\
Hi-Net~\cite{zhou2020hi} & T1+DWI$\to$T2 &87.64 $\pm$ 2.90 & 26.58 $\pm$ 1.18 &0.0659 $\pm$ 0.030 &1.532 $\pm$ 0.447 \\
MMgSN-Net~\cite{li2022virtual} & T1+DWI$\to$T2 &89.70 $\pm$ 2.95 & 27.40 $\pm$ 1.26 &0.0628 $\pm$ 0.032 &1.436 $\pm$ 0.463 \\
\underline{IMPORTANT-Net} & \underline{T1+DWI$\to$T2} &\underline{91.23 $\pm$ 3.09} & \underline{29.02 $\pm$ 1.38} &\underline{0.0597 $\pm$ 0.029} &\underline{1.085 $\pm$ 0.395} \\
\textbf{IMPORTANT-Net} & \textbf{T1+DWI+DCE$\to$T2} &\textbf{91.82 $\pm$ 3.15} & \textbf{30.75 $\pm$ 1.51} &\textbf{0.0531 $\pm$ 0.028} &\textbf{0.978 $\pm$ 0.397} \\
 \hline
\vspace{-15pt}
\end{tabular}
\end{table*}

In addition, the PSNR of IMPORTANT-Net (T1+DWI+DCE) is 30.75 $\pm$ 1.51, which is higher than all other cohort models, which indicates that the images synthesized by IMPORTANT-Net are of higher quality. At the same time, IMPORTANT-Net has the lowest NMSE of 0.0531 $\pm$ 0.028, which indicates that the error between the synthetic image and real image of IMPORTANT-Net is lower than that of other cohort models. In particular, to evaluate the model's performance on the synthesis of deep features, we compare the perceptual losses of different models. The synthetic image and the real image were used as inputs, respectively, and the outputs of the pre-trained VGG19 were obtained to calculate the perceptual loss (Equation \ref{eq:LP}). As shown in Table~\ref{table1}, IMPORTANT-Net has the lowest $L_P$ of 0.978 $\pm$ 0.397, which indicates that our proposed model can not only achieve higher overall performance, but also efficiently synthesize deep features of images.

\subsection{Visualization} \label{sec:viz}
\begin{figure}[htbp]
	\centering
	\includegraphics[width=1\linewidth]{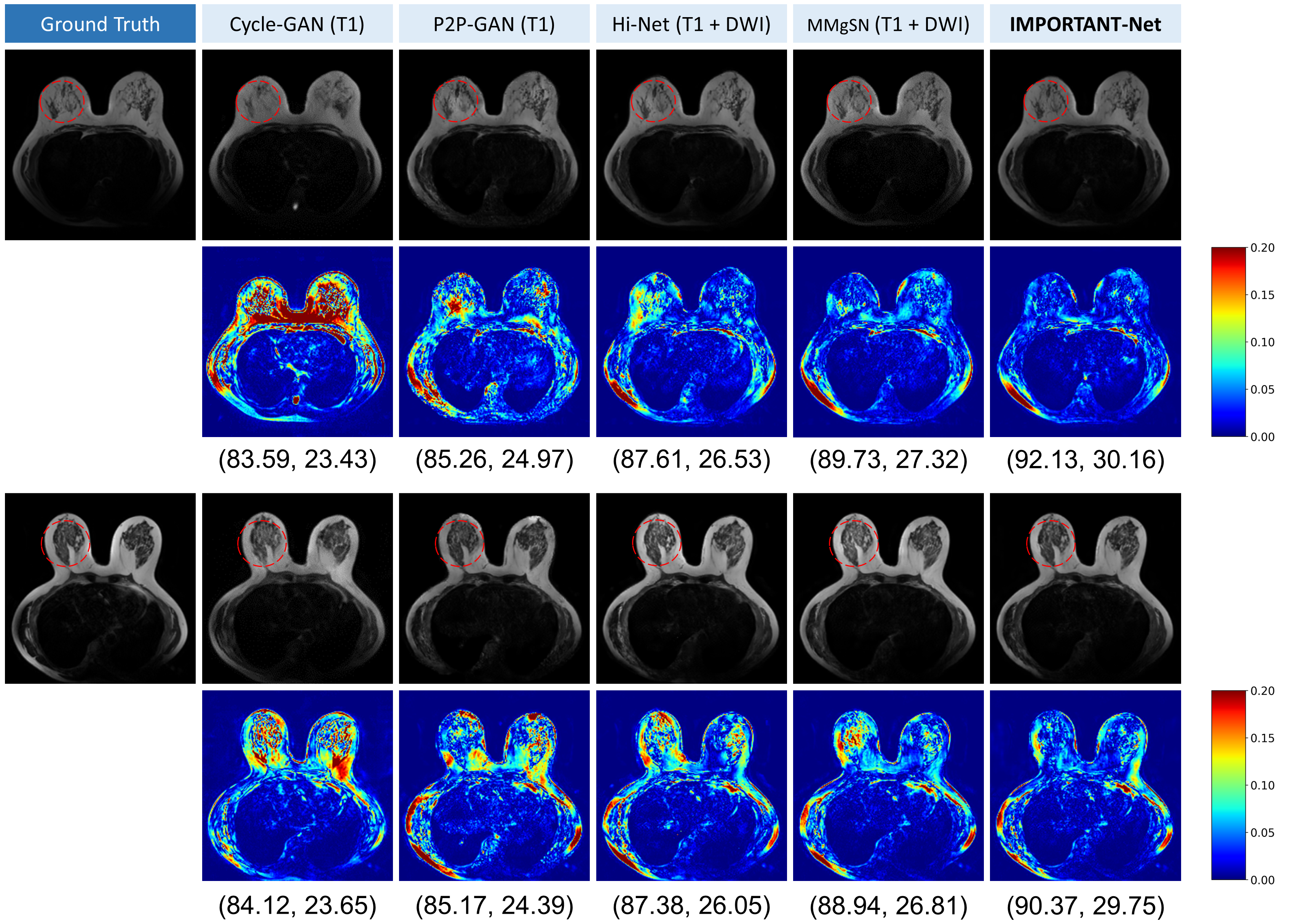}
	\caption{The visualization of the synthesized images using different models. (SSIM, PSNR) values are shown at the bottom.}
    \label{fig:fig4}
\end{figure}

In addition to the quantitative comparisons described above, intuitive qualitative results were also used to evaluate the models. Here, we directly compare the performance of the models by visualizing the error maps of different models. Some examples are shown in Fig.~\ref{fig:fig4}~(a), where the synthetic images based on Cycle-GAN had the largest error compared to the real images, which is evident from the higher rate of red areas (large error values, meaning highly irrelevant) than in the other models. Although the highly irrelevant regions of P2P-GAN and Hi-Net are reduced compared with Cycle-GAN, the effect is still unsatisfactory. In contrast, the red and yellow regions of the error maps of the images synthesized by our proposed model IMPORTANT-Net were significantly reduced, indicating that our model outperforms the cohort models in qualitative results. In addition, the visualization results from different views of the images are also used for comparison (Please see Supplementary Materials). 

\vspace{-5pt}

\begin{table*}[htb]
\centering
\caption{Ablation results for synthesizing T2-weighted breast MRI on the in-house dataset.}
\label{table2}
\setlength{\tabcolsep}{5pt}
\begin{tabular}{llcccc}
\hline 
Method & Parameter & SSIM$\uparrow$ & PSNR$\uparrow$ & NMSE$\downarrow$ & $L_{P}$$\downarrow$ \\ \hline
MP-net & T1+DWI+DCE$\to$T2 &85.65 $\pm$ 2.95 & 25.57 $\pm$ 1.12 &0.1271 $\pm$ 0.025 &1.806 $\pm$ 0.481 \\
MPF-net & T1+DWI+DCE$\to$T2 &89.95 $\pm$ 3.06 &27.86 $\pm$ 1.39 &0.0752 $\pm$ 0.027 &1.412 $\pm$ 0.472 \\
MPFA-net & T1+DWI+DCE$\to$T2 &91.55 $\pm$ 3.11 &29.51 $\pm$ 1.43 &0.0606 $\pm$ 0.033 &1.263 $\pm$ 0.390\\
\textbf{IMPORTANT-Net} & \textbf{T1+DWI+DCE$\to$T2} &\textbf{91.82 $\pm$ 3.15} & \textbf{30.75 $\pm$ 1.51} &\textbf{0.0531 $\pm$ 0.028} &\textbf{0.978 $\pm$ 0.397} \\
 \hline
\vspace{-20pt}
\end{tabular}
\end{table*}

\begin{table*}[htb]
\centering
\caption{Results for synthesizing T1-weighted breast MRI for different models on the in-house dataset.}
\label{table3}
\setlength{\tabcolsep}{5pt}
\begin{tabular}{llcccc}
\hline 
Method & Parameter & SSIM$\uparrow$ & PSNR$\uparrow$ & NMSE$\downarrow$ & $L_{P}$$\downarrow$ \\ \hline
Cycle-GAN~\cite{zhu2017unpaired} & T2$\to$T1 &83.15 $\pm$ 3.73 &23.41 $\pm$ 2.35 &0.1722 $\pm$ 0.038 &2.859 $\pm$ 0.526 \\
P2P-GAN~\cite{isola2017image} & T2$\to$T1 &85.11 $\pm$ 2.60 &25.01 $\pm$ 1.29 &0.1307 $\pm$ 0.029 &2.051 $\pm$ 0.509 \\
Hi-Net~\cite{zhou2020hi} & T2+DWI$\to$T1 &87.03 $\pm$ 2.65 &26.50 $\pm$ 1.68 &0.0729 $\pm$ 0.031 &1.695 $\pm$ 0.455 \\
MMgSN-Net~\cite{li2022virtual} & T2+DWI$\to$T1 &88.92 $\pm$ 2.69 &27.12 $\pm$ 1.79 &0.0681 $\pm$ 0.030 &1.506 $\pm$ 0.473 \\
\textbf{IMPORTANT-Net} & \textbf{T2+DWI$\to$T1} &\textbf{90.15 $\pm$ 2.92} &\textbf{28.51 $\pm$ 1.83} &\textbf{0.0602 $\pm$ 0.032} &\textbf{1.393 $\pm$ 0.470} \\
 \hline
\vspace{-10pt}
\end{tabular}
\end{table*}

\begin{table*}[htb]
\centering
\caption{Results for synthesizing T2-weighted brain MRI for different models on the BraTS2021 dataset.}
\label{table4}
\setlength{\tabcolsep}{5pt}
\begin{tabular}{llcccc}
\hline 
Method & Parameter & SSIM$\uparrow$ & PSNR$\uparrow$ & NMSE$\downarrow$ & $L_{P}$$\downarrow$ \\ \hline
Cycle-GAN~\cite{zhu2017unpaired} & T1$\to$T2 &86.83 $\pm$ 2.97 &27.82 $\pm$ 1.75 &0.1061 $\pm$ 0.029&2.126 $\pm$ 0.501 \\
P2P-GAN~\cite{isola2017image} & T1$\to$T2 &89.92 $\pm$ 2.48 &29.08 $\pm$ 1.07 &0.0777 $\pm$ 0.010&1.747 $\pm$ 0.457 \\
Hi-Net~\cite{zhou2020hi} & T1+Flair$\to$T2 &91.88 $\pm$ 2.06 &30.17 $\pm$ 1.37 &0.0712 $\pm$ 0.016&1.457 $\pm$ 0.404 \\
MMgSN-Net~\cite{li2022virtual} & T1+Flair$\to$T2 &92.55 $\pm$ 1.82 &30.73 $\pm$ 1.39 &0.0690 $\pm$ 0.017&1.226 $\pm$ 0.393 \\
\underline{IMPORTANT-Net} & \underline{T1+Flair$\to$T2} &\underline{94.07 $\pm$ 1.41} &\underline{32.06 $\pm$ 1.49} &\underline{0.0543 $\pm$ 0.009}&\underline{0.895 $\pm$ 0.308} \\
\textbf{IMPORTANT-Net} & \textbf{T1+Flair+T1ce$\to$T2} & \textbf{94.50 $\pm$ 1.45} &\textbf{32.51 $\pm$ 1.46} &\textbf{0.0527 $\pm$ 0.010}& \textbf{0.847 $\pm$ 0.317}\\
 \hline
\vspace{-20pt}
\end{tabular}
\end{table*}
\vspace{-10pt}
\subsection{Ablation study} \label{sec:as}
As described in Methods, the proposed model IMPORTANT-Net consists of several key components, including a parameter reconstruction module, MAPS fusion strategy, multi-parameter attention module, hierarchical $V^-$-shape generation module, etc. Therefore, ablation studies were performed to demonstrate the importance and effectiveness of our three key components. Several network structures were selected for comparison, as follows: (1) Multi-parameter synthesis network without parameter reconstruction module, MAPS fusion strategy and $V^-$-shape synthesis module (called MP-net), (2) Multi-parameter synthesis network with parameter reconstruction module and MAPS fusion strategy (called MPF-net), (3) MPF combined with parameter attention module (MPFA-net), and (4) IMPORTANT-Net reinforcement with $V^-$-shape generation module.
As shown in Table~\ref{table2}, MPF-net performed significantly better than MP-net, SSIM improved from 85.65 ± 2.95 to 89.95 $\pm$ 3.06, and PSNR improved from 25.57 $\pm$ 1.12 to 27.86 $\pm$ 1.39. This indicates that the parameter reconstruction module can effectively extract the latent representation information of the multi-parameter MRI images, and the fusion process of the MAPS strategy enables the information between the multi-parameter MRIs to interact through the "interaction room" to effectively synthesize the target image. In addition, the introduction of the parameter attention module has further improved the performance of the model, with SSIM of 91.55 $\pm$ 3.11, PSNR of 29.51 $\pm$ 1.43, NMSE of 0.0606 $\pm$ 0.033, and $L_P$ of 1.263 $\pm$ 0.390. This may be because the parameter attention module gives different weights to the representation information of different MRI parameters, enhancing valuable information and weakening useless information, further improving the model performance.
Finally, the hierarchical $V^-$-shape reinforcement generation module was constructed by skip-connecting the MPFA module to form the final IMPORTANT-Net. Compared with the queue model, IMPORTANT-Net achieved the best performance, with SSIM of 91.82 $\pm$ 3.15, PSNR of 30.75 $\pm$ 1.51, NMSE of 0.0531 $\pm$ 0.028, and $L_P$ of 0.978 $\pm$ 0.397. This is likely due to the representation information in the "analysis path" that was obtained by the parameter reconstruction module, and these representation information was greatly compressed during the encoding process, which would cause difficulties in the "synthesis path" (such as the up-sampling process needs to fill in lots of blank content). In contrast, the hierarchical $V^-$-shape reinforcement generation module here introduces the feature information at the corresponding scale into up-sampling through skip connections, providing multi-scale and multi-level information for the "synthesis path", thereby improving the performance of the model.
\vspace{-5pt}
\subsection{Extension study and discussion} \label{sec:esd}
To verify the model's versatility for coping with various inputs, we also evaluated the model's performance on synthetic T1-weighted images. As shown in Table~\ref{table3}, the IMPORTANT-Net model performed the best with a SSIM of 90.15$\pm$ 2.92, a PSNR of 28.51 $\pm$ 1.83, a NMSE of 0.0602 $\pm$ 0.032, and a $L_P$ of 1.393 $\pm$ 0.470, outperforming other cohort models including Cycle-GAN, P2P-GAN, Hi-Net and MMgSN-Net. This further demonstrates the excellent ability of our proposed IMPORTANT-Net in synthesizing missing/absent MRI data. In addition, as shown in Table~\ref{table3}, we specifically compare the performance of IMPORTANT-Net and Hi-Net in synthesizing T1-weighted MRI based on two MRI parameters, and the results show that our IMPORTANT-Net also outperforms Hi-Net.

\begin{figure}[htbp]
	\centering
	\includegraphics[width=1\linewidth]{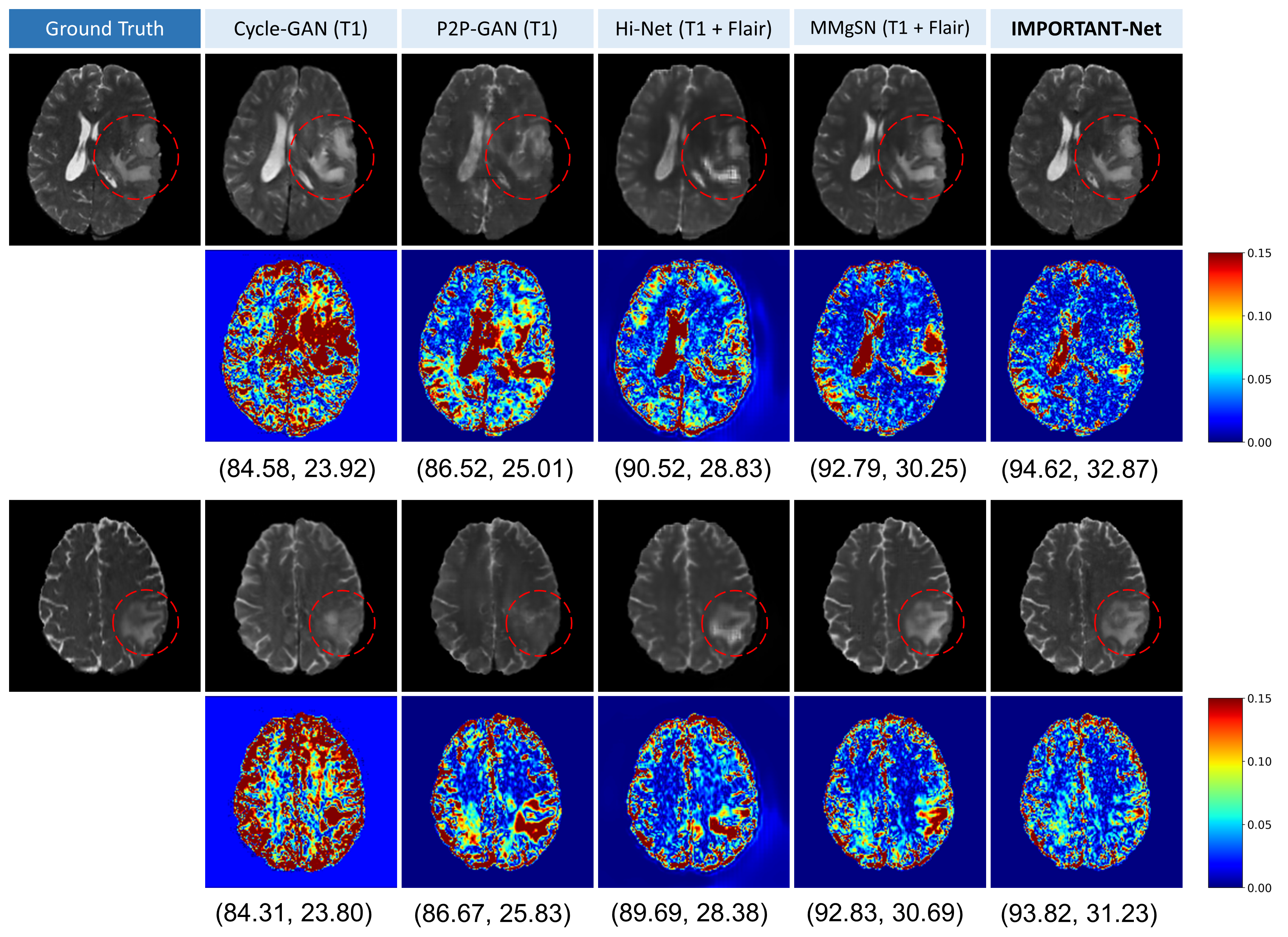}
	\caption{The visualization of the synthesized images using different models on the BraTS2021 dataset. (SSIM, PSNR) values are shown at the bottom.}
    \label{fig:fig5}
    \vspace{-20pt}
\end{figure}

Furthermore, we also validate the effectiveness of IMPORTANT-Net in synthesizing missing/absent data on another publicly available brain MRI dataset from Brain Tumor Segmentation 2021 (BraTS2021)~\cite{menze2014multimodal,baid2021rsna}. As shown in Table~\ref{table4}, our IMPORTANT-Net outperforms state-of-the-art methods. Some visualization examples are shown in Fig.~\ref{fig:fig5}. The red and yellow regions of the error maps of the images synthesized by IMPORTANT-Net are significantly reduced, indicating that our model outperforms the cohort models in qualitative results. See Supplementary Materials for visualizations from different views and other results.

Currently, existing methods, such as Cycle-GAN~\cite{zhu2017unpaired} and P2P-GAN~\cite{isola2017image}, mostly only focus on generating another image from one type of image, i.e. a single input and output, such as from MRI to CT, from CT to MRI, from T1 to T2, from T2 to T1, etc. However, these methods do not contribute to synthesizing target images by combination of different parameters from multiple images. Although some recent studies combine multiple images~\cite{zhou2020hi,li2022virtual}, the effect is still unsatisfactory. In contrast, as shown Fig.~\ref{fig:fig1}, our model can combine any number of existing parameters/modalities to synthesize missing/absent data, which will benefit the full utilization and mining of existing data. Furthermore, the simple concatenation of existing models for multi-modality/multi-parameter data in the input phase cannot effectively learn the underlying relationships of different information. In this study, in response to existing problems, the parameter reconstruction module is designed to extract the representation information from different parameters, and the multi-parameter fusion module is specially designed to enable the representation information from multiple parameters to be interactively fused through the MAPS strategic interaction room. In particular, the parameter attention module is designed to optimize the weights of different information, which further enhances the performance of the model. Regarding deep learning, it is well known that large amounts of data are critical to the success of training, and multi-parameter multi-modal data have also been shown to improve model performance in medical imaging-related tasks, such as tumor segmentation tasks and lesion classification tasks~\cite{dalmis2019artificial,hu2020deep,conte2021generative}. In addition, multi-parameter data can also provide more reference for clinicians. Therefore, IMPORTANT-Net has great potential and application value to contribute to these tasks and scenarios, to enrich data types, and to provide additional reference for clinicians. In clinical practice this could be used to benefit abbreviated MRI protocols. The demonstrated generalizability and versatility of our model in synthesizing missing/absent data implies that our method can be used on diverse input from various MRI protocols.
\section{Conclusion}
\label{sec:conclusion}
\vspace{-5pt}
In this paper, we have developed the IMPORTANT-Net to fully exploit multi-parameter information and fuse latent representations reinforced at different scale level for synthesizing missing/absent data on our in-house dataset (breast MRIs) and the public BraTS2021 dataset (brain MRIs). The method outperformed state-of-arts. Parametric reconstruction module, multi-parameter fusion with attention module and hierarchical $V^-$-shape reinforcement generation module are all proven to lead to cumulative improvement for synthesizing target images. The provided technical innovations for medical image synthesis can potentially benefit healthcare in the use scenarios when certain MRI sequences are absent and it may also be applied for other modality-image synthesis.

\section*{Acknowledgments}
The authors thank to the support from the Oversea Study Program of Guangzhou Elite Project. .

\bibliographystyle{unsrt}  
\bibliography{references}

\begin{thebibliography}{10}

\bibitem{sung2021global}
Hyuna Sung, Jacques Ferlay, Rebecca~L Siegel, Mathieu Laversanne, Isabelle
  Soerjomataram, Ahmedin Jemal, and Freddie Bray.
\newblock Global cancer statistics 2020: Globocan estimates of incidence and
  mortality worldwide for 36 cancers in 185 countries.
\newblock {\em CA: a cancer journal for clinicians}, 71(3):209--249, 2021.

\bibitem{goldhirsch2013personalizing}
Aron Goldhirsch, Eric~P Winer, AS~Coates, RD~Gelber, Martine Piccart-Gebhart,
  B~Th{\"u}rlimann, H-J Senn, Kathy~S Albain, Fabrice Andr{\'e}, Jonas Bergh,
  et~al.
\newblock Personalizing the treatment of women with early breast cancer:
  highlights of the st gallen international expert consensus on the primary
  therapy of early breast cancer 2013.
\newblock {\em Annals of oncology}, 24(9):2206--2223, 2013.

\bibitem{mann2019breast}
Ritse~M Mann, Nariya Cho, and Linda Moy.
\newblock Breast mri: State of the art.
\newblock {\em Radiology}, 292(3):520--536, 2019.

\bibitem{van2021factors}
Kay~JJ van~der Hoogt, Robert~J Schipper, Gonneke~A Winter-Warnars, Leon~C
  Ter~Beek, Claudette~E Loo, Ritse~M Mann, and Regina~GH Beets-Tan.
\newblock Factors affecting the value of diffusion-weighted imaging for
  identifying breast cancer patients with pathological complete response on
  neoadjuvant systemic therapy: a systematic review.
\newblock {\em Insights into imaging}, 12(1):1--22, 2021.

\bibitem{dalmis2019artificial}
Mehmet~U Dalmis, Albert Gubern-M{\'e}rida, Suzan Vreemann, Peter Bult, Nico
  Karssemeijer, Ritse Mann, and Jonas Teuwen.
\newblock Artificial intelligence--based classification of breast lesions
  imaged with a multiparametric breast mri protocol with ultrafast dce-mri, t2,
  and dwi.
\newblock {\em Investigative radiology}, 54(6):325--332, 2019.

\bibitem{hu2020deep}
Qiyuan Hu, Heather~M Whitney, and Maryellen~L Giger.
\newblock A deep learning methodology for improved breast cancer diagnosis
  using multiparametric mri.
\newblock {\em Scientific reports}, 10(1):1--11, 2020.

\bibitem{conte2021generative}
Gian~Marco Conte, Alexander~D Weston, David~C Vogelsang, Kenneth~A Philbrick,
  Jason~C Cai, Maurizio Barbera, Francesco Sanvito, Daniel~H Lachance, Robert~B
  Jenkins, W~Oliver Tobin, et~al.
\newblock Generative adversarial networks to synthesize missing t1 and flair
  mri sequences for use in a multisequence brain tumor segmentation model.
\newblock {\em Radiology}, 299(2):313--323, 2021.

\bibitem{dar2019image}
Salman~UH Dar, Mahmut Yurt, Levent Karacan, Aykut Erdem, Erkut Erdem, and Tolga
  Cukur.
\newblock Image synthesis in multi-contrast mri with conditional generative
  adversarial networks.
\newblock {\em IEEE transactions on medical imaging}, 38(10):2375--2388, 2019.

\bibitem{zhou2020hi}
Tao Zhou, Huazhu Fu, Geng Chen, Jianbing Shen, and Ling Shao.
\newblock Hi-net: hybrid-fusion network for multi-modal mr image synthesis.
\newblock {\em IEEE transactions on medical imaging}, 39(9):2772--2781, 2020.

\bibitem{gommers2021breast}
Jessie~JJ Gommers, Adri~C Voogd, Mireille~JM Broeders, Vivian van
  Breest~Smallenburg, Luc~JA Strobbe, Astrid~B Donkers-van Rossum, Hermen~C van
  Beek, Ritse~M Mann, and Lucien~EM Duijm.
\newblock Breast magnetic resonance imaging as a problem solving tool in women
  recalled at biennial screening mammography: A population-based study in the
  netherlands.
\newblock {\em The Breast}, 60:279--286, 2021.

\bibitem{sanderink2021diffusion}
Wendelien~BG Sanderink, Jonas Teuwen, Linda Appelman, Linda Moy, Laura Heacock,
  Elisabeth Weiland, Ioannis Sechopoulos, and Ritse~M Mann.
\newblock Diffusion weighted imaging for evaluation of breast lesions:
  Comparison between high b-value single-shot and routine readout-segmented
  sequences at 3 t.
\newblock {\em Magnetic resonance imaging}, 84:35--40, 2021.

\bibitem{geuzinge2021cost}
H~Amarens Geuzinge, Marije~F Bakker, Eveline~AM Heijnsdijk, Nicolien~T van
  Ravesteyn, Wouter~B Veldhuis, Ruud~M Pijnappel, St{\'e}phanie~V de~Lange,
  Marleen~J Emaus, Ritse~M Mann, Evelyn~M Monninkhof, et~al.
\newblock Cost-effectiveness of magnetic resonance imaging screening for women
  with extremely dense breast tissue.
\newblock {\em JNCI: Journal of the National Cancer Institute},
  113(11):1476--1483, 2021.

\bibitem{veenhuizen2021supplemental}
Stefanie~GA Veenhuizen, St{\'e}phanie~V de~Lange, Marije~F Bakker, Ruud~M
  Pijnappel, Ritse~M Mann, Evelyn~M Monninkhof, Marleen~J Emaus, Petra~K
  de~Koekkoek-Doll, Robertus~HC Bisschops, Marc~BI Lobbes, et~al.
\newblock Supplemental breast mri for women with extremely dense breasts:
  results of the second screening round of the dense trial.
\newblock {\em Radiology}, 299(2):278--286, 2021.

\bibitem{mann2022breast}
Ritse~M Mann, Alexandra Athanasiou, Pascal~AT Baltzer, Julia Camps-Herrero,
  Paola Clauser, Eva~M Fallenberg, Gabor Forrai, Michael~H Fuchsj{\"a}ger,
  Thomas~H Helbich, Fleur Killburn-Toppin, et~al.
\newblock Breast cancer screening in women with extremely dense breasts
  recommendations of the european society of breast imaging (eusobi).
\newblock {\em European Radiology}, pages 1--10, 2022.

\bibitem{mango2015abbreviated}
Victoria~L Mango, Elizabeth~A Morris, D~David Dershaw, Andrea Abramson, Charles
  Fry, Chaya~S Moskowitz, Mary Hughes, Jennifer Kaplan, and Maxine~S Jochelson.
\newblock Abbreviated protocol for breast mri: are multiple sequences needed
  for cancer detection?
\newblock {\em European journal of radiology}, 84(1):65--70, 2015.

\bibitem{chen2017abbreviated}
Shuang-Qing Chen, Min Huang, Yu-Ying Shen, Chen-Lu Liu, and Chuan-Xiao Xu.
\newblock Abbreviated mri protocols for detecting breast cancer in women with
  dense breasts.
\newblock {\em Korean journal of radiology}, 18(3):470--475, 2017.

\bibitem{leithner2019abbreviated}
Doris Leithner, Linda Moy, Elizabeth~A Morris, Maria~A Marino, Thomas~H
  Helbich, and Katja Pinker.
\newblock Abbreviated mri of the breast: does it provide value?
\newblock {\em Journal of Magnetic Resonance Imaging}, 49(7):e85--e100, 2019.

\bibitem{kim2022abbreviated}
Soo-Yeon Kim, Nariya Cho, Hyunsook Hong, Youkyoung Lee, Heera Yoen, Yeon~Soo
  Kim, Ah~Reum Park, Su~Min Ha, Su~Hyun Lee, Jung~Min Chang, et~al.
\newblock Abbreviated screening mri for women with a history of breast cancer:
  Comparison with full-protocol breast mri.
\newblock {\em Radiology}, 305(1):36--45, 2022.

\bibitem{heacock2016evaluation}
Laura Heacock, Amy~N Melsaether, Samantha~L Heller, Yiming Gao, Kristine~M
  Pysarenko, James~S Babb, Sungheon~G Kim, and Linda Moy.
\newblock Evaluation of a known breast cancer using an abbreviated breast mri
  protocol: correlation of imaging characteristics and pathology with lesion
  detection and conspicuity.
\newblock {\em European journal of radiology}, 85(4):815--823, 2016.

\bibitem{naranjo2022mri}
Isaac~Daimiel Naranjo, Julie Sogani, Carolina Saccarelli, Joao~V Horvat,
  Varadan Sevilimedu, Mary~C Hughes, Roberto~Lo Gullo, Maxine~S Jochelson,
  Jeffrey Reiner, and Katja Pinker.
\newblock Mri screening of brca mutation carriers: comparison of standard
  protocol and abbreviated protocols with and without t2-weighted images.
\newblock {\em American Journal of Roentgenology}, 218(5):810--820, 2022.

\bibitem{goodfellow2020generative}
Ian Goodfellow, Jean Pouget-Abadie, Mehdi Mirza, Bing Xu, David Warde-Farley,
  Sherjil Ozair, Aaron Courville, and Yoshua Bengio.
\newblock Generative adversarial networks.
\newblock {\em Communications of the ACM}, 63(11):139--144, 2020.

\bibitem{creswell2018generative}
Antonia Creswell, Tom White, Vincent Dumoulin, Kai Arulkumaran, Biswa Sengupta,
  and Anil~A Bharath.
\newblock Generative adversarial networks: An overview.
\newblock {\em IEEE signal processing magazine}, 35(1):53--65, 2018.

\bibitem{karras2019style}
Tero Karras, Samuli Laine, and Timo Aila.
\newblock A style-based generator architecture for generative adversarial
  networks.
\newblock In {\em Proceedings of the IEEE/CVF conference on computer vision and
  pattern recognition}, pages 4401--4410, 2019.

\bibitem{gui2021review}
Jie Gui, Zhenan Sun, Yonggang Wen, Dacheng Tao, and Jieping Ye.
\newblock A review on generative adversarial networks: Algorithms, theory, and
  applications.
\newblock {\em IEEE Transactions on Knowledge and Data Engineering}, 2021.

\bibitem{yi2019generative}
Xin Yi, Ekta Walia, and Paul Babyn.
\newblock Generative adversarial network in medical imaging: A review.
\newblock {\em Medical image analysis}, 58:101552, 2019.

\bibitem{zhu2017unpaired}
Jun-Yan Zhu, Taesung Park, Phillip Isola, and Alexei~A Efros.
\newblock Unpaired image-to-image translation using cycle-consistent
  adversarial networks.
\newblock In {\em Proceedings of the IEEE international conference on computer
  vision}, pages 2223--2232, 2017.

\bibitem{isola2017image}
Phillip Isola, Jun-Yan Zhu, Tinghui Zhou, and Alexei~A Efros.
\newblock Image-to-image translation with conditional adversarial networks.
\newblock In {\em Proceedings of the IEEE conference on computer vision and
  pattern recognition}, pages 1125--1134, 2017.

\bibitem{li2022virtual}
Wen Li, Haonan Xiao, Tian Li, Ge~Ren, Saikit Lam, Xinzhi Teng, Chenyang Liu,
  Jiang Zhang, Francis Kar-ho Lee, Kwok-hung Au, et~al.
\newblock Virtual contrast-enhanced magnetic resonance images synthesis for
  patients with nasopharyngeal carcinoma using multimodality-guided synergistic
  neural network.
\newblock {\em International Journal of Radiation Oncology* Biology* Physics},
  112(4):1033--1044, 2022.

\bibitem{menze2014multimodal}
Bjoern~H Menze, Andras Jakab, Stefan Bauer, Jayashree Kalpathy-Cramer, Keyvan
  Farahani, Justin Kirby, Yuliya Burren, Nicole Porz, Johannes Slotboom, Roland
  Wiest, et~al.
\newblock The multimodal brain tumor image segmentation benchmark (brats).
\newblock {\em IEEE transactions on medical imaging}, 34(10):1993--2024, 2014.

\bibitem{baid2021rsna}
Ujjwal Baid, Satyam Ghodasara, Suyash Mohan, Michel Bilello, Evan Calabrese,
  Errol Colak, Keyvan Farahani, Jayashree Kalpathy-Cramer, Felipe~C Kitamura,
  Sarthak Pati, et~al.
\newblock The rsna-asnr-miccai brats 2021 benchmark on brain tumor segmentation
  and radiogenomic classification.
\newblock {\em arXiv preprint arXiv:2107.02314}, 2021.

\bibitem{armanious2020medgan}
Karim Armanious, Chenming Jiang, Marc Fischer, Thomas K{\"u}stner, Tobias Hepp,
  Konstantin Nikolaou, Sergios Gatidis, and Bin Yang.
\newblock Medgan: Medical image translation using gans.
\newblock {\em Computerized medical imaging and graphics}, 79:101684, 2020.

\bibitem{welander2018generative}
Per Welander, Simon Karlsson, and Anders Eklund.
\newblock Generative adversarial networks for image-to-image translation on
  multi-contrast mr images-a comparison of cyclegan and unit.
\newblock {\em arXiv preprint arXiv:1806.07777}, 2018.

\bibitem{liu2019susan}
Fang Liu.
\newblock Susan: segment unannotated image structure using adversarial network.
\newblock {\em Magnetic resonance in medicine}, 81(5):3330--3345, 2019.

\bibitem{yu2019ea}
Biting Yu, Luping Zhou, Lei Wang, Yinghuan Shi, Jurgen Fripp, and Pierrick
  Bourgeat.
\newblock Ea-gans: edge-aware generative adversarial networks for
  cross-modality mr image synthesis.
\newblock {\em IEEE transactions on medical imaging}, 38(7):1750--1762, 2019.

\bibitem{yurt2021mustgan}
Mahmut Yurt, Salman~UH Dar, Aykut Erdem, Erkut Erdem, Kader~K Oguz, and Tolga
  {\c{C}}ukur.
\newblock mustgan: multi-stream generative adversarial networks for mr image
  synthesis.
\newblock {\em Medical image analysis}, 70:101944, 2021.

\bibitem{emami2018generating}
Hajar Emami, Ming Dong, Siamak~P Nejad-Davarani, and Carri~K Glide-Hurst.
\newblock Generating synthetic cts from magnetic resonance images using
  generative adversarial networks.
\newblock {\em Medical physics}, 45(8):3627--3636, 2018.

\bibitem{wolterink2017deep}
Jelmer~M Wolterink, Anna~M Dinkla, Mark~HF Savenije, Peter~R Seevinck,
  Cornelis~AT van~den Berg, and Ivana I{\v{s}}gum.
\newblock Deep mr to ct synthesis using unpaired data.
\newblock In {\em International workshop on simulation and synthesis in medical
  imaging}, pages 14--23. Springer, 2017.

\bibitem{jin2019deep}
Cheng-Bin Jin, Hakil Kim, Mingjie Liu, Wonmo Jung, Seongsu Joo, Eunsik Park,
  Young~Saem Ahn, In~Ho Han, Jae~Il Lee, and Xuenan Cui.
\newblock Deep ct to mr synthesis using paired and unpaired data.
\newblock {\em Sensors}, 19(10):2361, 2019.

\bibitem{maspero2018dose}
Matteo Maspero, Mark~HF Savenije, Anna~M Dinkla, Peter~R Seevinck, Martijn~PW
  Intven, Ina~M Jurgenliemk-Schulz, Linda~GW Kerkmeijer, and Cornelis~AT
  van~den Berg.
\newblock Dose evaluation of fast synthetic-ct generation using a generative
  adversarial network for general pelvis mr-only radiotherapy.
\newblock {\em Physics in Medicine \& Biology}, 63(18):185001, 2018.

\bibitem{johnson2016perceptual}
Justin Johnson, Alexandre Alahi, and Li~Fei-Fei.
\newblock Perceptual losses for real-time style transfer and super-resolution.
\newblock In {\em European conference on computer vision}, pages 694--711.
  Springer, 2016.

\bibitem{zhang2018unreasonable}
Richard Zhang, Phillip Isola, Alexei~A Efros, Eli Shechtman, and Oliver Wang.
\newblock The unreasonable effectiveness of deep features as a perceptual
  metric.
\newblock In {\em Proceedings of the IEEE conference on computer vision and
  pattern recognition}, pages 586--595, 2018.

\bibitem{avants2011reproducible}
Brian~B Avants, Nicholas~J Tustison, Gang Song, Philip~A Cook, Arno Klein, and
  James~C Gee.
\newblock A reproducible evaluation of ants similarity metric performance in
  brain image registration.
\newblock {\em Neuroimage}, 54(3):2033--2044, 2011.

\end{thebibliography}

\end{document}